\documentclass[twocolumn,showpacs,preprintnumbers,amsmath,amssymb,prl]{revtex4}
\usepackage{graphicx}
\usepackage{dcolumn}
\usepackage{bm}
\begin{document}
\title{Reply on the ``Comment on ``Investigating the phase diagram of
finite extensive and nonextensive systems'' by Al. H. Raduta and
Ad. R. Raduta'' by A. S. Botvina and I. N. Mishustin''}
\author{Al. H. Raduta$^{1,2}$, Ad. R. Raduta$^{1,2}$}
\affiliation{
  $^1$GSI, D-64220 Darmstadt, Germany\\
  $^2$NIPNE, RO-76900 Bucharest, Romania}

\maketitle

The Botvina and Mishustin's comment addresses a detail (minor subject
to the aim and topic of the paper, i.e. {\em the Coulomb effects on
  the liquid-gas phase transition}) of the recent paper \cite{prl},
namely a free volume approach aimed to allow some qualitative
estimation of the system's critical point.  In their view this
distorts the behaviour of the pressure ($P$) versus volume ($V$)
curves due to the ``non-conservation of the center of mass (c.m.) of
the system'' implied by the employed parametrization.

While we agree with the evident idea that, apart from the case of the
large systems, the center of mass of a small system should be
conserved in thermodynamical calculations in order to obtain accurate
results, one should remark that the Botvina and Mishustin's
argumentation is based on a specific view on freeze-out:
non-overlapping spherical fragments placed in a spherical
``recipient'' of volume $V$. The shape of the fragments is therefore
fixed for each fragmentation event. This is, indeed, a scenario widely
used in statistical models from the beginning of the nuclear
multifragmentation studies. Within such a scenario, the Botvina and
Mishustin's comments are surely correct and they apply to most of the
statistical calculations performed so far (MMMC model \cite{mmmc},
Koonin and Randrup model \cite{randrup}, Das Gupta and Mekjian
calculations \cite{dasgupta}, and even recent calculations performed
with the SMM model \cite{elliott}, etc.). Following this hypothesis, at a
given temperature, $P$ will converge to a finite value when
$V$ will approach the volume of the source nucleus at normal density
($V_0$) (see e.g. \cite{botvina}).

However, dynamical calculations clearly show that fragments are not at
all spherical at break-up but have various shapes variating from event
to event (see e.g. \cite{dinamica}). It means that the spherical
fragment hypothesis is actually {\em away} from reality and this has
to be reflected in all thermodynamical observables. In this case, the
system's position coordinates (entering the integration which leads to
the free volume term in the partition function) will be shared
between the positions of each fragment's c.m. and the {\em additional}
coordinates corresponding to each fragment {\em deformation}. (In other
words there are {\em more} than $N-1$ contributing degrees of freedom.)
Obviously, this detail will raise the power of $V_f$ to a value {\em
  larger} than the $N-1$ responsible for the finite value of $P$ when
$V \rightarrow V_0$, and therefore $P$ will diverge near this limit
leading to the {\em well known} van der Waals behavior (see for
example the nuclear phase diagrams from Ref. \cite{De}). This very
simple reasoning shows that the conclusions of Botvina and Mishustin
are actually in disagreement with what one would expect from a {\em
  real} nuclear system.

But let us now discuss the free volume approach we used
Ref. \cite{prl}. This approach was employed in Ref. \cite{prl} with the
intention of reaching densities $\rho \rightarrow \rho_0$ unaffordable
with a spherical fragment scenario. Within this scenario it is assumed
that each fragment is blocking for itself and the rest of the
fragments its own volume.  Obviously, at $\rho \rightarrow \rho_0$
such a formula can only be justified by a deformed fragment
scenario. And this differs a lot from what Botvina and Mishustin claim
to understand from our paper: {\em the removal of the physical
assumption that fragments don't overlap}. The latter assumption is
indeed unphysical while the deformed fragment scenario (actually used
in our calculations) {\em is not}.

Of course, the deformed fragment scenario deserves a more refined
treatment than our calculations from Ref. \cite{prl}: a proper account
for the fragments' deformation degree of freedom, a sharp conservation
of the center of mass, an accurate treatment of the binding energies,
etc. Definitely, these aspects have to be addressed in the statistical
models for a better description of the {\em real} multifragmentation.

Summarizing, the Botvina and Mishustin's $P(V)_T$ curves correspond
to a simplified scenario for the freeze-out: nonoverlapping spherical
fragments inside a spherical freeze-out recipient, which, as explained
before {\em differs} from the scenario corresponding to our free
volume parametrization: non-overlapping fragments with various shapes
at freeze-out. The latter scenario is {\em the realistic one}, being
in agreement with the results of the various dynamical simulations of
nuclear multifragmentation and, moreover, at $\rho \rightarrow \rho_0$
the fragments {\em have} to be deformed in order to fit the recipient.
As explained before, within this scenario the pressure {\em has} to
diverge while $V$ approaches $V_0$ (fact in agreement with our results
and in disagreement with Botvina and Mishustin ones) since the power
of $V_f$ will necessarily be greater than $N-1$.  More refined
treatments than ours are surely necessary for obtaining very accurate
results, but the qualitative character of our evaluation was already
stated in Ref. \cite{prl}.  It results that, in spite of the roughness
(commented above) of our approximation, the Botvina and Mishustin's
result [on the $P(V)_T$ curves] is in fact the ``misleading'' one
since it differs essentially from what is expected from {\em real}
systems.

\end{document}